\documentclass[5p,preprint]{elsarticle}
\biboptions{sort&compress}
\usepackage{lineno,isotope,mathrsfs}
\modulolinenumbers[5]
\usepackage{amsmath,amssymb,mathptmx,epsfig,bm,mathrsfs,feynmp}
\usepackage{color,slashed,nicefrac,exscale,multirow,soul}
\usepackage{times,txfonts,xcolor,graphicx,gensymb,tikz}
\usepackage[normalem]{ulem}
\usepackage[breaklinks,colorlinks,citecolor=blue]{hyperref}
\definecolor{red}{rgb}{0.8,0,0}
\definecolor{violet}{rgb}{0.4,0,0.4}
\definecolor{green}{rgb}{0,0.5,0.0}
\definecolor{navy}{rgb}{0.0,0.0,0.6}
\definecolor{orange}{rgb}{0.8,0.2,0.0}
\usepackage[normalem]{ulem} %\sout{old text} for strikeout  

\newcommand{\MR}{$M$-$R\ $}
\begin{document}
\begin{frontmatter}
\title{Baryonic models of ultra-low-mass compact stars for the \\
central compact object in HESS J1731-347}
\author[a]{Jia Jie Li}
\ead{jiajieli@swu.edu.cn}
\author[b,c]{Armen Sedrakian}
\ead{sedrakian@fias.uni-frankfurt.de}
\address[a]{School of Physical Science and Technology, Southwest University, Chongqing 400715, China}
\address[b]{Frankfurt Institute for Advanced Studies, D-60438
Frankfurt am Main, Germany}
\address[c]{Institute of Theoretical Physics, University of Wroclaw,
50-204 Wroclaw, Poland}
\begin{abstract}
The recent attempt on mass and radius inference of the central
compact object within the supernova remnant HESS J1731-347 suggests
for this object an unusually low mass of
$M = 0.77^{+0.20}_{-0.17}\,M_{\odot}$ and a small radius of
$R = 10.4^{+0.86}_{-0.78}$\,km. We explore the ways such a result
can be accommodated within models of dense matter with heavy
baryonic degrees of freedom which are constrained by the
multi-messenger observations. We find that to do so using only
purely nucleonic models, one needs to assume a rather small value of
the slope of symmetry energy $L_{\rm sym}$. Once heavy baryons are
included higher values of the slope $L_{\rm sym}$ become acceptable
at the cost of a slightly reduced maximum mass of static
configuration. These two scenarios are distinguished by the
particle composition and will undergo different cooling scenarios.
In addition, we show that the universalities of the $I$-Love-$Q$
relations for static configurations can be extended to very low
masses without loss in their accuracy.
\end{abstract}
\end{frontmatter}
%\linenumbers
%
%-------------------------------------------------------------------
\section{Introduction}
\label{sec:Intro}
%-------------------------------------------------------------------
Over the last decade, there has been exciting progress in gathering
new astrophysical information on compact star (CS) parameters such 
as mass, radius, and tidal deformability. This progress boosts the 
search for the ultimate state of dense matter and, simultaneously,
narrows down the admissible theories of dense matter.

The most massive neutron star yet observed, PSR J0740\-+6620, with a
mass of $2.08^{+0.07}_{-0.07}\,M_{\odot}$ (68.3\% CI)~\cite{NANOGrav:2019,Fonseca:2021},
excludes equation of state (EoS) models which cannot support star with 
masses that reach that limit. Recently the companion of ``black widow'' 
pulsar PSR J0952-0607 was reported to have a mass of
$2.35^{+0.17}_{-0.17}\,M_{\odot}$ (68.3\% CI)~\cite{Romani:2022}, which
makes it the fastest and heaviest known neutron star to date. The data 
from the NICER instrument and pulse modelling of X-ray emission 
from hot spots on the surfaces of CSs have provided simultaneous mass 
and radius measurements for PSR J0030+0451~\cite{NICER:2019a,NICER:2019b} 
and PSR J0740+6620~\cite{NICER:2021a,NICER:2021b}. The detection of 
gravitational wave (GW) event GW170817~\cite{LIGO_Virgo:2017} involving 
a merger of two CSs has placed constraints on the tidal deformability of
this system, $\tilde\Lambda \leq 720$~\cite{LIGO_Virgo:2018,LIGO_Virgo:2019,Desoumi:2018,Coughlin:2018,Kiuchi:2019}.
While the associated electromagnetic counterparts, AT2017gfo and
GRB170817A~\cite{Soares-Santos:2017,Villar:2017}, have provided an upper 
limit on the maximum static mass of neutron stars 
$M\sim 2.3\,M_{\odot}$~\cite{Shibata:2017,Margalit:2017,Ruiz:2018,Rezzolla:2018,Shibata:2019,Khadkikar:2021}.
The X-ray observations of the temperature of the neutron star in the
Cassiopeia A supernova remnant and modeling of its thermal evolution
give insights into the superfluid properties of the core
matter~(see Ref.~\cite{Shternin:2021} and references therein). 

Very recently, an estimate of the mass and radius of the central compact
object (CCO) within the supernova remnant HESS J1731-347 was
reported~\cite{Doroshenko:2022} using modeling of its X-ray spectrum and 
a distance estimate from Gaia observations. The mass and the radius of this 
object are $M = 0.77^{+0.20}_{-0.17}\,M_{\odot}$ and
$R = 10.4^{+0.86}_{-0.78}$\,km (68.3\% CI), respectively, when only 
parallax priors and X-ray data are considered. Taking in addition the 
full distance priors and EoS constraint priors into account, the improved 
estimates are $M = 0.83^{+0.17}_{-0.13}\,M_{\odot}$ and
$R = 11.25^{+0.53}_{-0.37}$\,km (68.3\% CI). This estimation makes the CCO 
in HESS J1731-347 the lightest neutron star known to 
date~\cite{Doroshenko:2022,Brodie:2023}, and potentially a candidate for an 
exotic object --- ``strange star''~\cite{Doroshenko:2022,Clemente:2022,Horvath:2023}, 
as stars made of ordinary matter in supernova explosions have a lower 
limit on the mass which is above $1.17\,M_{\odot}$~\cite{Suwa:2018}.
Previously, the confirmed lightest neutron star with
$M = 1.174^{+0.004}_{-0.004}\,M_{\odot}$ is the companion star in the
double neutron star system with the primary being the pulsar J0453+1559
which has an estimated mass~\cite{Martinez:2015,Ozel:2016}. Note
that these two mass estimates are compatible at better than 
2$\sigma$ accuracy.

Ref.~\cite{Doroshenko:2022} based their discussion on EoS models and
associated mass-radius ($M$-$R$) curves which were based on the $\chi$EFT 
EoS up to the density of $1.5\,\rho_{\rm sat}$ (where $\rho_{\rm sat}$ 
is the nuclear saturation density) that are extrapolated to higher
densities using the constant speed-of-sound EoS. It was shown that a
large number of EoS models meets all current observational
constraints. An alternative scheme of analysis is based on the 
covariant density functional (CDF) approach~\cite{Oertel:2017,Sedrakian:2023}, 
which is a powerful tool that allows one to model finite nuclei 
and nuclear matter across a wide range of densities and temperatures 
relevant to neutron stars and their coalescences. In this work, 
we test how well the CDF approach to baryonic matter can accommodate 
the putative mass and radius values inferred for the CCO in 
HESS J1731-347 by Ref.~\cite{Doroshenko:2022}.

%-------------------------------------------------------------------
\section{CDF for baryonic matter}
\label{sec:Model}
%-------------------------------------------------------------------
The theoretical framework implemented in this work has been presented 
in great detail in Ref.~\cite{Sedrakian:2023}. We provide below the most 
relevant points of our approach for the sake of completeness. The baryonic 
matter is described within a CDF approach with density-dependent 
baryon-meson couplings~\cite{Lalazissis:2005,Sedrakian:2023}. 
The interaction part of the Lagrangian is given by 
\begin{eqnarray}\label{eq:interaction_Lagrangian}
\mathscr{L}_{\text{int}}
&=& \sum_B \bar{\psi}_B\Big(-g_{\sigma B}\sigma-g_{\sigma^\ast B}\sigma^\ast
    -g_{\omega B}\gamma^\mu\omega_\mu-g_{\phi B}\gamma^\mu\phi_\mu \nonumber \\
&&  -g_{\rho B}\gamma^\mu\vec{\rho}_\mu\cdot\vec{\tau}_B\Big)\psi_B
    + \sum_D (\psi_B \rightarrow \psi^\nu_D),
\end{eqnarray}
where $\psi$ stands for the Dirac spinor with index $B$ labeling the 
spin-1/2 baryonic octet, namely, nucleons $N \in \{n,p\}$ and hyperons
$Y \in \{\Lambda,\Xi^{0,-},\Sigma^{+,0,-}\}$, while $\psi^\nu$ for the
Rarita-Schwinger spinor~\cite{Pascalutsa:2007} with index $D$ referring to 
the spin-3/2 resonance quartet $\Delta \in \{\Delta^{++,+,0,-}\}$.
The baryons interact via exchanges of $\sigma, \omega$ and $\rho$ mesons, 
which comprise the minimal set necessary for a quantitative description 
of nuclear phenomena~\cite{Serot:1997}. In addition, the two hidden-strangeness
mesons ($\sigma^\ast,\,\phi$) describe interactions between
hyperons~\cite{Schaffner:1994,Oertel:2015,Lijj:2018a}.

In the nucleonic sector, these are three meson-nucleon ($mN$) coupling
constants ($g_{\sigma N},\,g_{\omega N},\,g_{\rho N}$) at saturation 
density $\rho_{\rm sat}$, and four parameters that control their density 
dependence. These seven parameters allow one to establish the correspondence
between this type of a CDF and the purely phenomenological expansion of 
the energy density of nuclear matter~\cite{Margueron:2018,Lijj:2019b} in 
the vicinity of $\rho_{\rm sat}$, with respect to the number density $\rho$ 
and isospin asymmetry $\delta = (\rho_{\rm n}-\rho_{\rm p})/\rho$ where 
$\rho_{\rm n(p)}$ is the neutron (proton) density. The coefficients 
of this double expansion are referred to commonly as saturation energy 
$E_{\rm sat}$, incompressibility $K_{\rm sat}$, and the skewness $Q_{\rm sat}$ 
in isoscalar channel, and the symmetry energy $J_{\rm sym}$, and its 
slope parameter $L_{\rm sym}$ in isovector channel. 

Among the above-mentioned coefficients, of particular interest are the
quantities that arise at a higher order of the expansion, specifically, 
$Q_{\rm sat}$ and $L_{\rm sym}$. Their values are weakly constrained by 
the conventional fitting protocol used in constructing the density 
functionals. However, they play complementary roles as the value of 
$Q_{\rm sat}$ controls the high-density behavior of the nucleonic EoS, 
and thus the maximum mass of nucleonic CSs, while the value of $L_{\rm sym}$ 
determines the intermediate-density behavior of the nucleonic EoS, and 
thus the radii of low-mass stars.

In hyperonic sector, the vector meson-hyperon ($mY$) couplings are 
given by either the SU(6) spin-flavor symmetric quark model or the 
SU(3) flavor symmetric model~\cite{Schaffner:1994,Oertel:2015,Lijj:2018a,Sedrakian:2023}, 
depending on the stiffness of the utilized nucleonic EoS model. 
On the other hand the scalar meson-hyperon couplings are determined 
by fitting to certain preselected properties of hypernuclear systems. 
To be specific, as it is common, we determine the coupling constants, 
$g_{\sigma Y}$, using the following hyperon potentials in the symmetric 
nuclear matter at $\rho_{\rm sat}$~\cite{Feliciello:2015,Gal:2016}:
\begin{align}
U^{(\rm N)}_\Lambda = -U^{(\rm N)}_\Sigma = -30\,\text{MeV},\quad 
U^{(\rm N)}_\Xi = -14\,\text{MeV}.
\end{align}
In addition, we adopt the $\Lambda\Lambda$ bond energy at $\rho_{\rm{sat}}/5$,
which reproduces the most accurate experimental value to date~\cite{Ahnjk:2013},
\begin{align}
U^{(\Lambda)}_\Lambda  = - 0.67\,\text{MeV},
\end{align}
to fix the value of the coupling $g_{\sigma^\ast \Lambda}$.
The coupling of remaining hyperons $\Xi$ and $\Sigma$ to the $\sigma^*$
is determined by the relation
$g_{\sigma^\ast Y}/g_{\phi Y} = g_{\sigma^\ast\Lambda}/g_{\phi\Lambda}$~\cite{Lijj:2018a}.

Finally, let us turn to the $\Delta$-resonance sector.
The information on the meson-$\Delta$ ($m\Delta$) couplings is scarce,
as no consensus has been reached yet on the magnitude of the
$\Delta$ potential. In this work, we limit ourselves to the case where
\begin{align}
R_{\Delta\omega} = g_{\omega\Delta}/g_{\omega N} = 1.1, \quad 
R_{\Delta\rho} = g_{\rho\Delta}/ g_{\rho N} = 1.0,
\end{align}
and $R_{\Delta\sigma} = g_{\sigma\Delta}/g_{\sigma N}$
is determined by fitting to the following representative 
values of $\Delta$-potentials in nuclear matter at $\rho_{\rm sat}$,
\begin{align}
U^{(\rm N)}_\Delta = 1,\,4/3,\,5/3\,U_{\rm N},
\end{align}
where $U_{\rm N}$ is the nucleon isoscalar potential.
This means the potential of $\Delta$'s in the
nuclear medium is more attractive than the nucleon potential, 
consistent with the studies of the scattering of
electrons and pions off nuclei and photoabsorption with 
phenomenological models~\cite{Drago:2014}. The resultant 
values for $R_{\Delta\sigma}$, respectively, are, 
\begin{align}
R_{\Delta\sigma} \simeq 1.10,\,1.16,\,1.23.
\end{align}
Note that within the range
$(R_{\Delta\sigma} - R_{\Delta\omega})\gtrsim 0.15$, $\Delta$-admixed
matter undergoes spinodal instability~\cite{Raduta:2021}. Our choice
avoids such instabilities with the most attractive case
$U^{(\rm N)}_\Delta = 5/3\,U_{\rm N}$ being at the limit of the
stability. This completes our discussion of the parameters entering
our CDF for baryonic matter and they are now fully determined.
Note that we assume that the hyperon and $\Delta$ potentials 
scale with a density as the nucleonic one, therefore their 
high-density behavior is inferred from that of the nucleons.

With this input, we compute the EoS of the stellar matter by implementing 
the additional conditions of weak equilibrium and charge neutrality that 
prevail in CSs. We further match smoothly our EoS for the core to 
that of the crust EoS given in Refs.~\cite{Baym:1971a,Baym:1971b} at 
the crust-core transition density $\sim \rho_{\rm{sat}}/2$.

In the present work, as in Refs.~\cite{Lijj:2019a,Lijj:2019b}, we map
the nucleonic EoS given by the 
Lagrangian~\eqref{eq:interaction_Lagrangian} 
for each set of parameters $Q_{\rm sat}$ and $L_{\rm sym}$. 
For our analysis below we adopt the lower-order coefficients in the 
isoscalar channel, i.e., $E_{\rm sat} = -16.14$, and 
$K_{\rm sat} = 251.15$\,MeV, as those inferred from the DDME2 
parametrization~\cite{Lalazissis:2005} which were adjusted to the 
properties of finite nuclei. For the isovector channel, instead 
fixing $J_{\rm sym}$ = 32.31\,MeV the DDME2 value, we hold the value 
of the symmetry energy at a lower density $\rho_{\rm c} = 0.11$\,fm$^{-3}$ 
at the constant value $E_{\rm sym}(\rho_{\rm c}) = 27.09$\,MeV~\cite{Lijj:2019b}. 
The density $\rho_{\rm c}$ is distinguished by the fact that a large 
variety of nuclear models predict almost identical values of 
$E_{\rm sym}(\rho_{\rm c})$. In the following discussion, we will 
use the pair of parameters ($Q_{\rm sat}$,\,$L_{\rm sym}$) to 
identify any specific nucleonic CDF.

%---------------------------Nuclear-Matter--------------------------
\begin{figure}[tb]
\centering
\includegraphics[width = 0.42\textwidth]{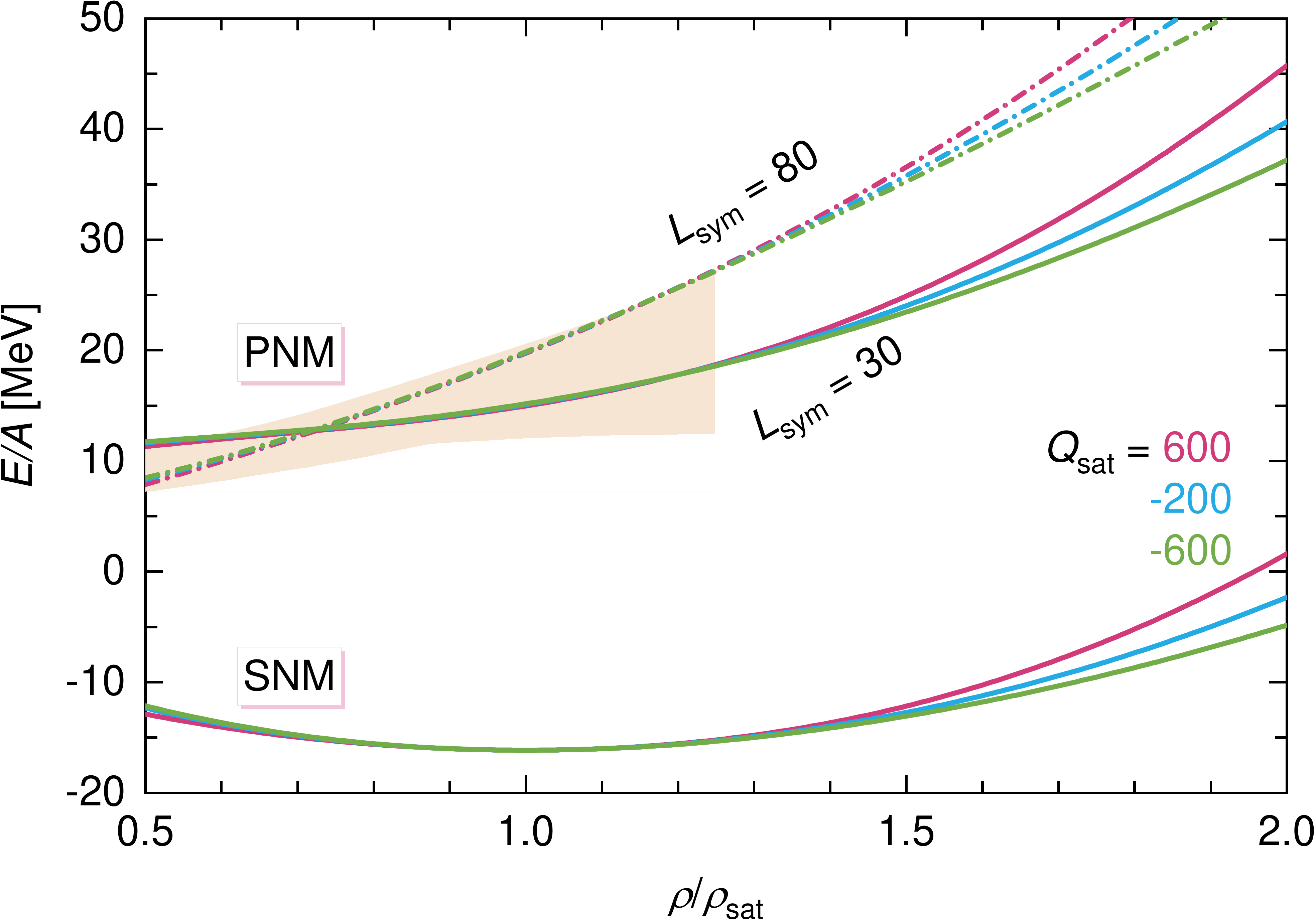}
\caption{Energy per particle of symmetric nucleonic matter (SNM) and
  pure neutron matter (PNM) as a function of density
  $\rho/\rho_{\rm sat}$, obtained from six representative
  ($Q_{\rm sat},\,L_{\rm sym}$) pairs (in MeV). The band corresponds
  to the combined $\chi$EFT results from Ref.~\cite{Huth:2021}.}
\label{fig:EOS_nucl}
\end{figure}
%-------------------------------------------------------------------

To appreciate the quality of current functional and the range of
variation of parameters we show in Fig.~\ref{fig:EOS_nucl} the 
energy per particle of symmetric nucleonic matter (SNM) and 
pure neutron matter (PNM) as a function of  density for six 
representative ($Q_{\rm sat},\,L_{\rm sym}$) pairs, while keeping 
all other coefficients at their default values of DDME2 
parametrization~\cite{Lalazissis:2005,Lijj:2019b}.
We further restrict the set of EoS by choosing only those which reproduce 
the combined result for PNM derived from several recent many-body 
calculations with $\chi$EFT interactions for densities up to 
$\sim 1.3\,\rho_{\rm sat}$~\cite{Huth:2021}. This results in the range 
$30 \leqslant L_{\rm sym} \leqslant 80$\,MeV, which is independent of 
the values of $Q_{\rm sat}$. In addition, we have checked that all 
the alternative parametrizations can reasonably reproduce the binding 
energies and charge radii of several closed-shell nuclei with 
$\sim 2\%$ relative deviation.

%-------------------------------------------------------------------
\section{Results and discussion}
\label{sec:Result}
%-------------------------------------------------------------------
%
\subsection{CSs with only nucleons}
We start our discussion with the CS models containing only
nucleons. We limit the discussion to three sets of representative EoS
models by taking three values of $Q_{\rm sat} = -600,\,-200$, and
600\,MeV, and a range of $L_{\rm sym}$ which allow an overlap of our
models with the range limited by the analysis of the CCO in HESS
J1731-347~\cite{Doroshenko:2022}. Large values of $L_{\rm sym}$
correspond to a stiffer EoS close to nuclear saturation density, as
shown in Fig.~\ref{fig:EOS_nucl}, and lead to larger radii for
low-mass stars. Larger values of $Q_{\rm sat}$ imply EoS which are
stiffer at high density and, thus, predict larger maximum masses for
static nucleonic stars~\cite{Lijj:2019b,Zhangnb:2018}. For
$Q_{\rm sat} = -600$\,MeV the maximum mass is about $2.0\,M_\odot$,
which matches the mass of PSR
J0740+6620~\cite{NANOGrav:2019,Fonseca:2021}; for
$Q_{\rm sat} = -200$\,MeV the maximum mass is consistent with the
(approximate) {\it theoretical upper limit} on the maximum mass of
static CSs $\sim 2.3\,M_\odot$ inferred from analysis of 
the GW170817 event and the corresponding electromagnetic 
counterparts in a scenario where a supramassive remnant formed after 
a merger collapses into a black hole~\cite{Shibata:2017,Margalit:2017,Ruiz:2018,Rezzolla:2018,Shibata:2019,Khadkikar:2021};
finally, for $Q_{\rm sat} = 600$\,MeV the maximum mass is slightly
higher than $2.5\,M_\odot$, which would be compatible with the mass of
the secondary in the GW190814 event~\cite{LIGO_Virgo:2020} and its
interpretation as a nucleonic
CS~\cite{Fattoyev:2020,Sedrakian:2020,Lijj:2020b}.

%---------------------------MR-relation-Nucleonic--------------------------
\begin{figure}[tb]
\centering
\includegraphics[width = 0.42\textwidth]{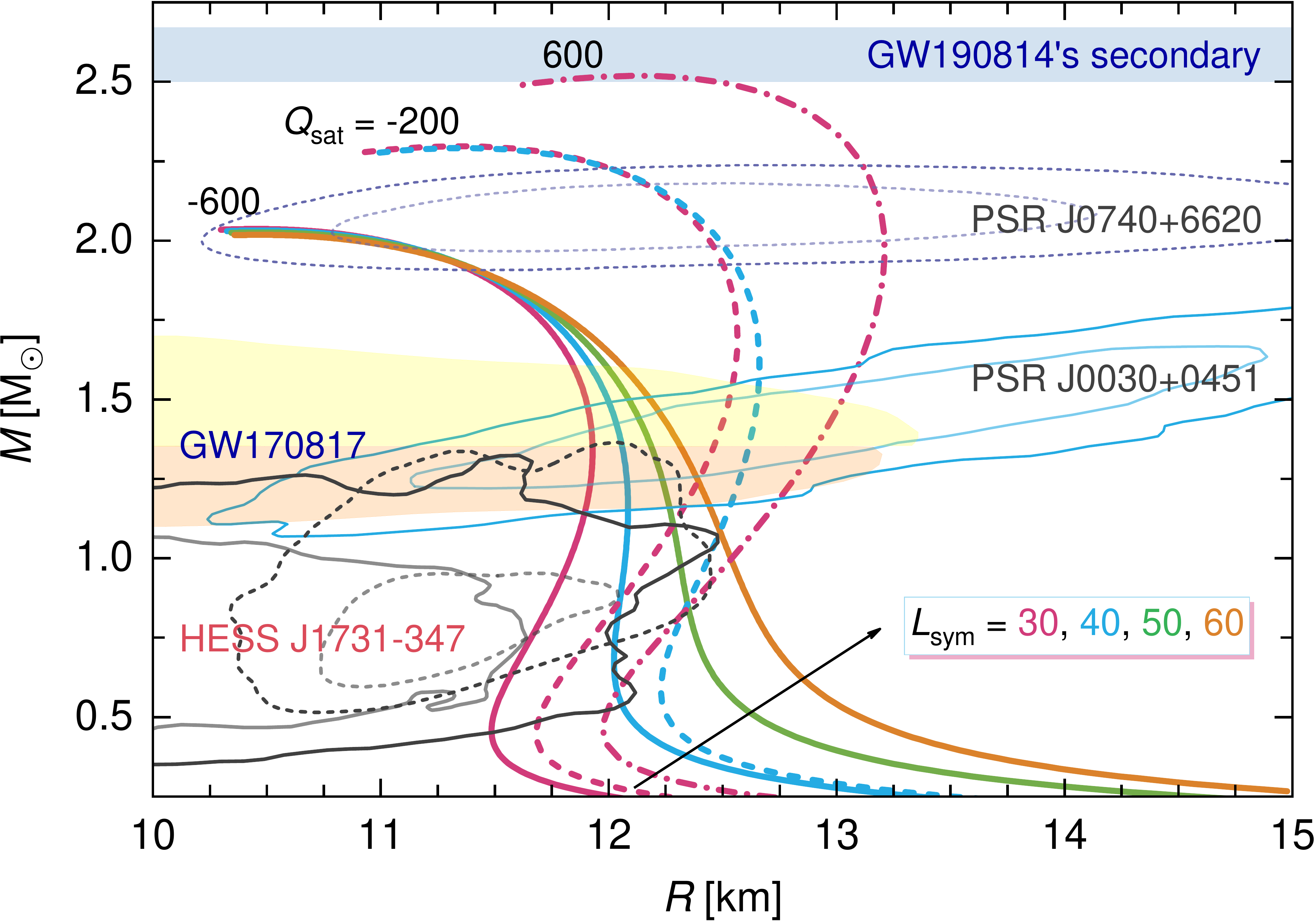}
\caption{\MR relation for nucleonic EoS models with different pairs of values
  of $Q_{\rm sat}$ and $L_{\rm sym}$. We show three branches of \MR
  curves, for $Q_{\rm sat}= -600$ (solid lines), $-200$ (dashed lines) and 
  $600$\,MeV (dash-doted lines). For each of these, $L_{\rm sym}$ is 
  varied from $30$\,MeV to larger values that are still compatible with 
  the ellipse of HESS J1731-347 at 95.4\% CI. The shaded regions show the 
  constraints from analysis of GW events~\cite{LIGO_Virgo:2018,LIGO_Virgo:2020}, 
  the ellipses indicate the regions compatible with the inferences from NICER
  observations~\cite{NICER:2019a,NICER:2019b,NICER:2021a,NICER:2021b},
  the contours show the \MR constraints for the CCO in HESS
  J1731-347~\cite{Doroshenko:2022}. See text for details.}
\label{fig:MR_Nucl}
\end{figure}
%------------------------------------------------------------------

The \MR relations for our nucleonic EoS models are shown in
Fig.~\ref{fig:MR_Nucl}, along with the current astrophysical
observational bands derived from X-ray or gravitational wave
observations. These include: (i) the ellipses obtained by the 
NICER collaboration which simultaneously determined the radius 
and mass of PSR J0030+0451 and J0740+6620 via X-ray pulse-profile 
modeling (at 68.3\% and 95.4\% CIs)~\citep{NICER:2019a,NICER:2019b,NICER:2021a,NICER:2021b};
(ii) the \MR contours for the CCO in HESS J1731-347, where solid 
lines correspond to the case that only parallax priors and X-ray 
data are considered, and the dashed lines correspond to the joint fit 
including all prior information (at 68.3\% and 95.4\%
CIs)~\cite{Doroshenko:2022}; (iii) the regions for each
of the two CSs that merged in the GW170817 event (at 90\%
CIs)~\citep{LIGO_Virgo:2019}; and (iv) the mass of the
secondary component of the GW190814 event (at 90\% CI)~\citep{LIGO_Virgo:2020}.

%-----------------------MR-relation-Heavy-baryons------------------
\begin{figure*}[tb]
\centering
\includegraphics[width = 0.92\textwidth]{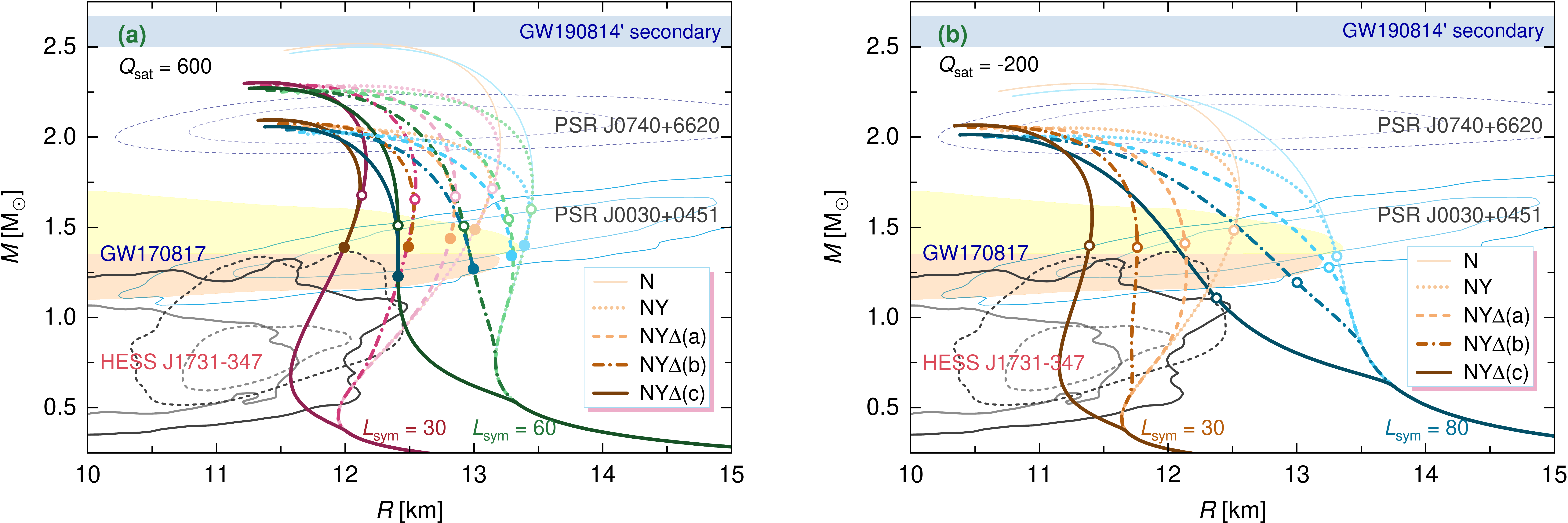}
\caption{\MR relation for hyperon-$\Delta$ admixed EoS models for
  different $\Delta$ potential depths at nuclear saturation density
  $U_\Delta/U_{\rm N} = 1,\,4/3,\,5/3$, which are labeled as
  ``NY$\Delta$(a)-(c)'', respectively. The results for purely
  nucleonic and hyperonic EoS models are also shown. In panel (a) the
  EoS models are constructed from the nucleonic model with pairs of
  $(Q_{\rm sat},\,L_{\rm sym})=(600,\,30)$ and $(600,\,60)$\,MeV, combined
  with either SU(6) or a SU(3) symmetric model for the hyperonic sector. In
  panel (b) the EoS models are constructed from the nucleonic model
  with pairs of $(Q_{\rm sat},\,L_{\rm sym})=(-200,\,30)$ and
  $(-200,\,80)$\,MeV and a SU(3) symmetric parametrization of the hyperonic
  sector. The onset mass of hyperons for each EoS model is marked by
  circles. See text for details.}
\label{fig:MR_Delt}
\end{figure*}
%--------------------------------------------------------------------

As seen from Fig.~\ref{fig:MR_Nucl}, these observational ellipses can
be accounted for by appropriate choices of the parameters $Q_{\rm sat}$
and $L_{\rm sym}$ at the 95.4\% CI. The choice of these parameters gives 
sufficient flexibility to allow for intermediate-density soft EoS to 
account for ellipses for CCO in HESS J1731-347 and GW170817 and sufficiently 
stiff EoS at high density to account for the ellipses of PSR~J0740+6620. 
To give more specifical examples, a 95.4\% CI consistency is achieved by choosing
$L_{\rm sym} \lesssim 60$\,MeV and $Q_{\rm sat} \sim -600$\,MeV,
$L_{\rm sym} \lesssim 40$\,MeV and $Q_{\rm sat} \sim -200$\,MeV, and
$L_{\rm sym} \lesssim 30$\,MeV and $Q_{\rm sat} \sim 600$\,MeV. 
Among these, the model with $(Q_{\rm sat},\,L_{\rm sym}) = (-600,\,30)$\,MeV 
intersects also the 68.3\% CI ellipse. Thus, we conclude that the nucleonic 
models are consistent with the current information available from 
multimessenger astrophysics within 95.4\% CI, in particular, the small 
radius of the CCO in HESS J1731-347, {\it if fairly low values of $L_{\rm sym}$ are adopted}. 
These low values for $L_{\rm sym}$ lie outside the 1$\sigma$ confidence placed by the
analysis of neutron skin in the PREX-II experiment for $^{208}$Pb~\cite{PREX-II:2021,Reed:2021},
but are compatible with the analysis including also dipole
polarizabilities in a set of finite nuclei~\cite{Reinhard:2021},
the analysis of the CREX experiment for $^{48}$Ca~\cite{CREX:2022,Reinhard:2022}, 
as well as with recent $\chi$EFT predictions~\cite{Drischler:2020}.

A large sample of nucleonic EoS based on CDF theory with 
nonlinear couplings  was considered in Ref.~\cite{Brodie:2023} and  
was found to be inconsistent with the 68.3\% CI ellipses of HESS 
J1731-347. This is in contrast to our finding, even though our EoS 
are consistent with the $\chi$EFT computations of PNM 
(see Fig.~\ref{fig:EOS_nucl}) which were used to tune the CDFs 
in Ref.~\cite{Brodie:2023}. The difference in the conclusions arises 
from the fact that our collection includes EoS with small values of 
the slope of symmetry energy, e.g., 
$(Q_{\rm sat},\,L_{\rm sym}) \simeq (-600,\,30)$\,MeV, which are 
still consistent with the $\chi$EFT band. This is in contrast, the 
collection of Ref.~\cite{Brodie:2023} which is limited to larger values 
of $L_{\rm sym} \simeq 40$\,MeV. Anticipating the discussion in the next 
section, we point out here that if the $\Delta$-resonance threshold is 
low enough, their nucleation relieves further the tension between the 
observations and CDF-based models.

\subsection{CSs with heavy baryons}
Next, we consider models which allow for the nucleation of heavy baryons in
dense matter. The hyperonic EoS models that support massive two-solar
mass CSs require the nucleonic sector to be a priori stiff,
which results in a large radius of a canonical-mass star
$R_{1.4} \gtrsim 13$\,km~\cite{Weissenborn:2012a,Weissenborn:2012b,Colucci:2013,Oertel:2015,Tolos:2016,Fortin:2017,Lijj:2018a}.
Only during recent years, hyperon-$\Delta$ admixed CDF-based models
were developed which were compatible with the nuclear data, radius measurements 
of CSs, and the tidal deformability inferred from the analysis of the GW170817
event~\cite{Lijj:2018b,Lijj:2019a,Ribes:2019,Lijj:2020a,Sedrakian:2023}. 
The appearance of $\Delta$-resonances, in parallel with hyperons, 
does not affect the maximum mass of a static CS, but reduces the radius 
of the star by tens of percent, depending on the value of $\Delta$-potential 
in nuclear matter $U_\Delta$. Therefore, it is convenient for further discussion to
denote underlying EoS model by the potential of $\Delta$'s, whereby the 
parameters specifying the nucleonic and hyperonic sectors are fixed.

In Fig.~\ref{fig:MR_Delt} we present the \MR relations for purely nucleonic, 
hyperonic, and hyperon-$\Delta$ admixed stellar matter for several parameter 
values, along with current astrophysical observational constraints. In panel 
(a) the EoS models are constructed from nucleonic model with $Q_{\rm sat} = 600$\,MeV, 
SU(6) symmetric parameterization for the hyperonic sector which results in a maximum mass 
$M_{\rm max} \simeq 2.0\,M_{\odot}$, and a SU(3) symmetric parameterization which 
yields $M_{\rm max} \simeq 2.3\,M_{\odot}$. To assess the range of 
variations in the sequences arising from the uncertainties in $\Delta$ 
potential $U_\Delta$ and the value of $L_{\rm sym}$ in nucleonic sector
which are crucial for the radius of the star, we consider for each model 
$U_\Delta/U_{\rm N}= 1,\,4/3,$ and 5/3 for $L_{\rm sym}= 30,\,60$\,MeV. 
In panel (b) we present results for alternative EoS models which are constructed 
from nucleonic model has $Q_{\rm sat} = -200$\,MeV in combination with a SU(3) 
symmetric model for the hyperonic sector, which gives $M_{\rm max} \simeq 2.0\,M_{\odot}$.

%----------------------------Particle Fraction-------------------
\begin{figure*}[tb]
\centering
\includegraphics[width = 0.92\textwidth]{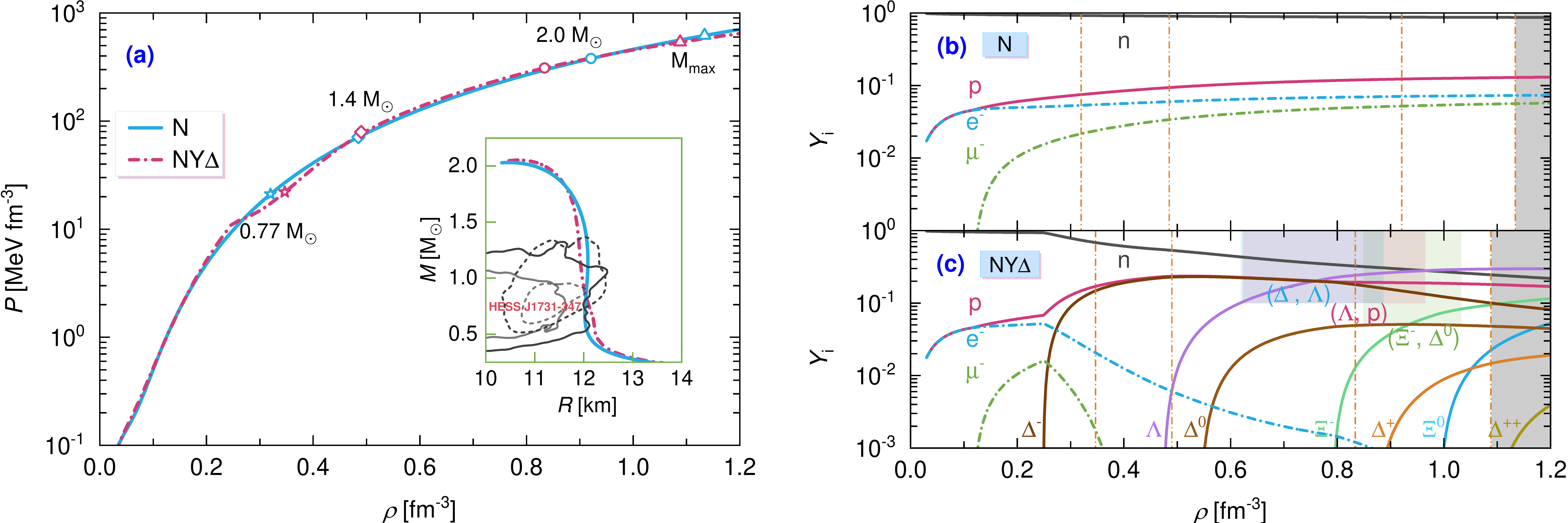}
\caption{EoS and the corresponding particle fraction for nucleonic
  and hyperon-$\Delta$ admixed models that support a $0.77\,M_{\odot}$
  CS with radius $\sim 12.0$\,km and a maximum mass of
  $\sim 2.0\,M_{\odot}$. The nucleonic EoS corresponds to the pair
  $(Q_{\rm sat},\,L_{\rm sym})=(-600,\,40)$\,MeV, while the
  hyperon-$\Delta$ admixed EoS is constructed from the nucleonic model
  with $(Q_{\rm sat},\,L_{\rm sym})=(-200,\,40)$\,MeV, combined with a
  SU(3) symmetric parametrization of the hyperonic sector and
  $U_\Delta/U_{\rm N}= 5/3$ value of $\Delta$ potential. In the left
  panel the positions of the respective 0.77,\,1.4,\,$2.0\,M_{\odot}$
  CSs and the maximum-mass configurations are marked by
  different symbols, while in the right panel, the corresponding
  central densities are indicated by thick vertical lines. In the
  right panel the gray shaded area shows the densities beyond the
  maximum-mass configurations. For hyperon-$\Delta$ admixed EoS
  model the regions of density where direct Urca processes are allowed
  are indicated by shadings with different colors.}
\label{fig:Fraction}
\end{figure*}
%----------------------------------------------------------------

In Fig.~\ref{fig:MR_Delt}\,(a), we show the results for the nucleonic
model with $(Q_{\rm sat},\,L_{\rm sym})=(600,\,30)$~MeV which are marginally
compatible with the CCO in HESS J1731-347 constraint at 95.4\% CI
limit. Allowing for hyperons and $\Delta$-resonances and varying the 
value of $U_\Delta$ we observe the reduction of the stellar radius with 
increasing value of $U_\Delta$. The \MR track in this case could lie within 
the 68.3\% CI ellipse. For a (isovector) stiffer nucleonic EoS model with
$(Q_{\rm sat},\,L_{\rm sym})=(600,\,60)$\,MeV which produces \MR tracks
outside the HESS J1731-347 region, the onset of hyperons and resonances 
leads to \MR tracks that are within 95.4\% CI region if a sufficiently 
attractive $\Delta$ potential is chosen, e.g., $U_\Delta/U_{\rm N} = 5/3$. 
In Fig.~\ref{fig:MR_Delt}\,(b), for EoS models that are constructed from a 
(isoscalar) softer nucleonic model with $Q_{\rm sat} = -200$\,MeV, the 
allowed range for $L_{\rm sym}$ is enlarged from $L_{\rm sym}\lesssim 40$\,MeV 
in purely nucleonic models to $L_{\rm sym} \lesssim 80$\,MeV in heavy baryons 
admixed models. We thus conclude that the tight constraint on the value of 
$L_{\rm sym}$ set by requirement of their compatibility with HESS J1731-347, 
is relaxed by about 30-40\,MeV if nucleation of heavy baryons in dense matter 
is allowed. Even though such models help to reconcile the realistic values of
$L_{\rm sym}$ with low-mass putative observations, it should be noted
that if HESS J1731-347 is a baryonic star it contains likely only
non-strange particles, due to its low mass. Indeed, as shown
in Fig.~\ref{fig:MR_Delt}, the onset of hyperons occurs in stars with
larger mass with $M \gtrsim 1.2\,M_{\odot}$. 

\subsection{Composition of the CCO in HESS J1731-347}
Fig.~\ref{fig:Fraction} shows the EoS and composition for nucleonic
and heavy baryon admixed models that support a $0.77\,M_{\odot}$ CS
with radius $\sim 12.0$\,km and a maximum mass $\sim 2.0\,M_{\odot}$.
The nucleonic EoS has $(Q_{\rm sat},\,L_{\rm sym})=(-600,\,40)$\,MeV;
while the $\Delta$-admixed hyperonic EoS model has
$(Q_{\rm sat},\,L_{\rm sym})=(-200,\,40)$\,MeV, a SU(3) symmetric
parameterization in the hyperonic sector and
$U_\Delta/U_{\rm N}= 4/3$. These two EoS models yield similar \MR
relations which are shown in the inset of Fig.~\ref{fig:Fraction}\,(a), and
the deviation in radius for a fixed mass is found less than 3\%.
For star with a mass $\sim 0.77\,M_{\odot}$, the inner core density is about 
$2\,\rho_{\rm sat}$, around where significant difference in particle 
composition is observed. Thus, the knowledge of the mass and radius 
alone does not allow one to distinguish the internal composition of CSs.
In fact, even under the restrictive assumption of purely 
nucleonic stars, current empirical knowledge allows for multiple 
solutions  for the composition of $\beta$-equilibrium
matter, i.e., the composition of CSs cannot be determined 
unequivocally~\cite{Mondal:2022}.

However, as well known, the particle content of dense matter is an 
important factor in the cooling of CSs due to new channels
of neutrino emissions in the presence of heavy baryons.
Because of same values of $L_{\rm sym}$ the nucleonic and
heavy baryon admixed models lead to a similar composition at low density 
$\rho \lesssim 0.25$\,fm$^{-3}$ in Figs.~\ref{fig:Fraction}\,(b) and (c). 
We note that the requirement of a small radius of CCO in HESS J1731-347,
and equivalently a low value of coefficient $L_{\rm sym}$ preclude the
onset of the direct Urca process in such models. In contrast, in the
$\Delta$-admixed hyperonic matter a trace fraction of hyperons and
$\Delta$'s will open up a host of direct Urca processes involving
these particles.  According to Fig.~\ref{fig:Fraction}\,(c) the
$\Delta$-admixed hyperonic EoS allows the direct Urca processes in a
large region of the density, $0.62$-$1.03$\,fm$^{-3}$, specifically
$\Delta^- \rightarrow \Lambda + e^- + \bar{\nu}_e$,
$\Lambda \rightarrow p + e^- + \bar{\nu}_e$, and
$ \Xi^- \rightarrow \Delta^0 + e^- + \bar{\nu}_e$, are allowed and
their thresholds are indicated in Fig.~\ref{fig:Fraction}\,(c) as well.
The onset of hyperons then will affect the cooling behaviour of stars
more massive than HESS J1731-347, see Ref.~\cite{Sedrakian:2023} and
references therein.

\subsection{Tidal deformability and $I$-Love-$Q$ relations}
%----------------------------M-TD relation---------------------
\begin{figure}[tb]
\centering
\includegraphics[width = 0.42\textwidth]{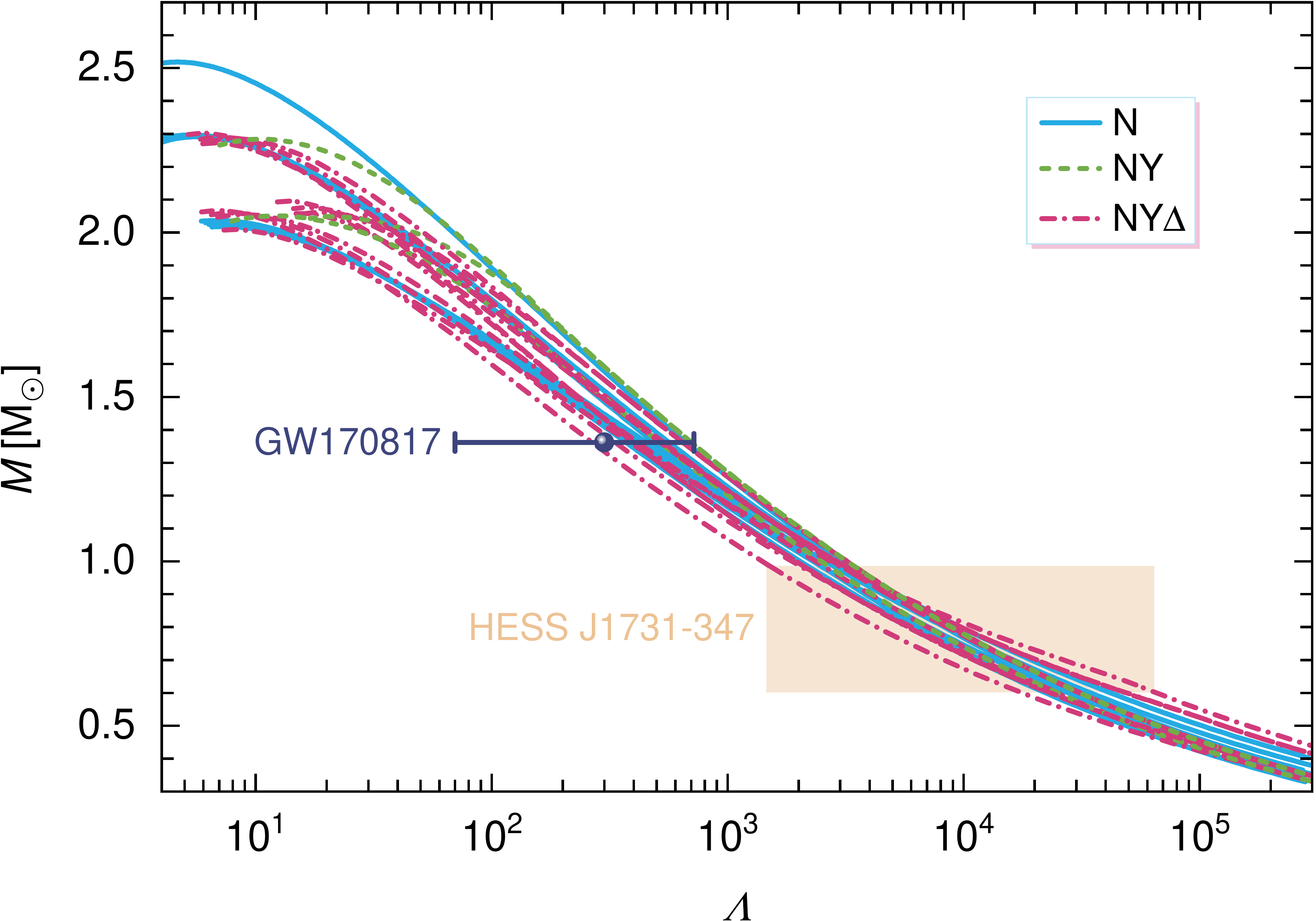}
\caption{Mass-tidal deformability relation for various EoS models and
  matter compositions. The results are derived from a collection of
  our models that satisfy all relevant astrophysical constraints at
  95.4\% CI. These include nucleonic ($N$), hyperonic ($NY$), and 
  hyperon-$\Delta$ admixed ($NY\Delta$) one. The constraint for a
  $1.36\,M_{\odot}$ star deduced from the analysis of 
  GW170817~\cite{LIGO_Virgo:2018,LIGO_Virgo:2019} is shown for comparison.  
  The mass range for a star with $M = 0.77^{+0.20}_{-0.17}\,M_{\odot}$ corresponding 
  to the inference for CCO in HESS J1731-347~\cite{Doroshenko:2022} is 
  indicated by shading.}
\label{fig:Mass_lamb}
\end{figure}
%--------------------------------------------------------------

%----------------------------I-Love-Q--------------------------
\begin{figure*}[tb]
\centering
\includegraphics[width = 0.92\textwidth]{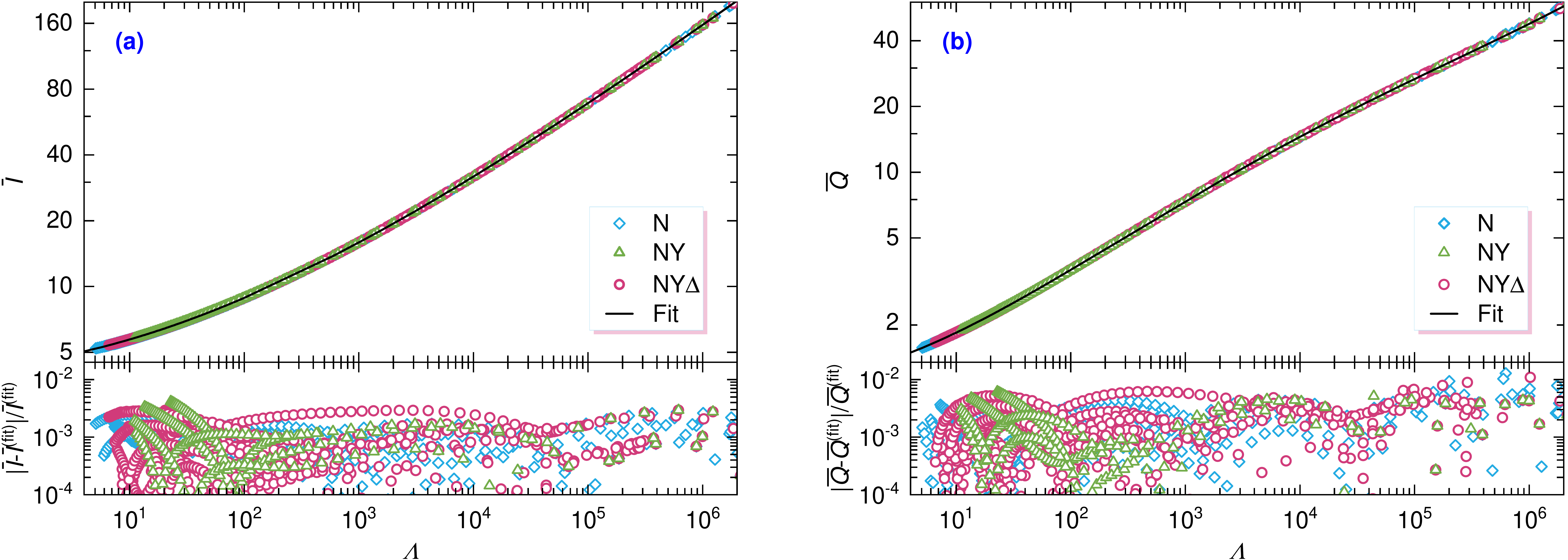}
\caption{The $I$-Love and $Q$-Love relations for nonrotating CSs. 
The top panels show universal relations for various EoS models and 
matter compositions, together with fitting curves; bottom panels show 
fractional errors between the fitting curves and numerical results.
The results are derived from a collection of valid models
that satisfy all relevant astrophysical constraints at the 95.4\%
CI for the cases of nucleonic ($N$), hyperonic ($NY$), and 
hyperon-$\Delta$ admixed ($NY\Delta$) compositions.}
\label{fig:ILQ_relation}
\end{figure*}
%---------------------------------------------------------------

The gross quantities of a CS such as the mass, radius, deformability, 
moment of inertia, quadrupole moment, etc., sensitively depend on the 
microscopic EoS. In Fig.~\ref{fig:Mass_lamb} we show the mass vs 
dimensionless tidal deformability relations for our models that satisfy all 
relevant astrophysical constraints at the 95.4\% CI. All these models satisfy 
the constraint placed for a $1.36\,M_{\odot}$ star from the analysis 
of the  GW170817 event. The ranges for the estimated mass 
$M = 0.77^{+0.20}_{-0.17}\,M_{\odot}$ of CCO in HESS J1731-347 are indicated as well. 
Since the low-mass stars have very large tidal deformability $\Lambda$, spanning 
the range from $10^3$ to $10^5$, it is clear that they could be good 
targets for GW observatories if involved in a merger process.

Various approximately universal relations connecting different CS
properties have been established and intensively studied in recent years.
Of particular interest is the $I$-Love-$Q$ relations that first discovered in 
Ref.~\cite{Yagi:2013} which connect the dimensionless moment of inertia 
$\bar{I}$, tidal deformability $\Lambda$, and the spin-induced quadrupole
moment $\bar{Q}$ of neutron stars in slow rotation approximation.
These relations can be numerically described with 
a polynomial~\cite{Yagi:2013,Yagi:2017}:
\begin{align}\label{Eq:ILQ_relation}
\ln y = a_0+a_1\ln x+a_2\,(\ln x)^2 + a_3\,(\ln x)^3+a_4\,(\ln x)^4,
\end{align}
where pairs $(x,\,y)$ represent $(\Lambda,\,\bar{I})$, $(\Lambda,\,\bar{Q})$ 
and $(\bar{Q},\,\bar{I})$. These relations are commonly studied for stars with 
mass $M \gtrsim 1.0\,M_{\odot}$~\cite{Yagi:2017}.

%------------------------------------------------------------------------
\begin{table}[tb]	
\centering
\caption{
The coefficients of the fit formulas of the 
$I$-Love-$Q$ relations.
}
\setlength{\tabcolsep}{4.6pt}
\begin{tabular}{llll}		
\hline\hline
      & $\bar{I}$-$\Lambda$       & $\bar{Q}$-$\Lambda$      & $\bar{I}$-$\bar{Q}$      \\ 
\hline	
$a_0$ & $ 1.48978 \times 10^{ 0}$ & $ 1.84748\times 10^{-1}$ & $ 1.28033\times 10^{0}$  \\                                          
$a_1$ & $ 6.45216 \times 10^{-2}$ & $ 1.01466\times 10^{-1}$ & $ 8.81910\times 10^{-1}$ \\            
$a_2$ & $ 2.12766 \times 10^{-2}$ & $ 4.48923\times 10^{-2}$ & $-2.92930\times 10^{-1}$ \\  
$a_3$ & $-5.95664 \times 10^{-4}$ & $-3.87686\times 10^{-3}$ & $ 1.42509\times 10^{-1}$ \\   
$a_4$ & $ 5.15851 \times 10^{-6}$ & $ 1.07983\times 10^{-4}$ & $-1.56774\times 10^{-2}$ \\                                          
\hline\hline       
\end{tabular}
\label{table-IloveQ}
\end{table}
%------------------------------------------------------------------------\\

Here we briefly investigate whether these relations hold 
for stars with heavy baryons, in particular, the hyperon-$\Delta$
admixed CSs, especially in the low-mass region extending down 
to about $0.2\,M_{\odot}$. Our results for $I$-Love and $Q$-Love 
relations, derived from a collection of our models that satisfy 
all relevant astrophysical constraints at the 95.4\% CI, 
are shown in Fig.~\ref{fig:ILQ_relation}, together with the fits 
according to Eq.~\eqref{Eq:ILQ_relation}, where the bottom panels 
present the fractional differences between the data and the fits.
It is seen in Fig.~\ref{fig:ILQ_relation} that the absolute fractional 
differences are $\lesssim 1\%$ for all two relations. Clearly, the 
universality holds for the third pair $\bar{I}$-$\bar{Q}$ as well. 
The best-fit coefficients with Eq.~\eqref{Eq:ILQ_relation} for the 
three relations are summarized in Table~\ref{table-IloveQ}, which are 
consistent with those found in Ref.~\cite{Yagi:2017}. Thus, our results 
extend the previously reported universal $I$-Love-$Q$ relations for 
CSs into the low-mass domain.

%----------------------------------------------------------------------
\section{Summary}
\label{sec:Summary}
%----------------------------------------------------------------------
In this work, we have studied the problem of the nature of the CCO
in HESS J1731-347 by using a CDF approach for dense matter with and 
without heavy baryons in high-density regions. While some authors suggested 
the ``strange star'' scenario for this object~\cite{Clemente:2022,Horvath:2023}, 
here we explore a ``less strange'' alternative. The density functional 
we used was parameterized and varied via two parameters - the skewness 
coefficient $Q_{\rm sat}$ of symmetric nuclear matter and the slope 
coefficient $L_{\rm sym}$ of the symmetry energy. The hyperon potentials 
were tuned to the most plausible potentials extracted from hypernuclear 
data. The $\Delta$-potential was taken to be in the range
$1 \leq U_\Delta/U_N \leq 5/3$, --- a reasonable range consistent with
the current data.

We found that purely nucleonic models for the EoS can accommodate the
estimate for mass and radius of the CCO in HESS J1731-347, but only if the
slope of symmetry energy coefficient $L_{\rm sym}$ is fairly small,
i.e., $L_{\rm sym} \lesssim 60$\,MeV for models with
$Q_{\rm sat} \sim -600$\,MeV, and $L_{\rm sym} \lesssim 30$\,MeV for
models with $Q_{\rm sat} \sim 600$\,MeV. These low values for
$L_{\rm sym}$ are compatible with the analysis of the PREX-II and CREX
experiments of neutron skins, dipole polarizabilities, as well as recent
$\chi$EFT computations. Allowing for hyperon and $\Delta$-resonance
populations in dense matter, our EoS models allow a broader range of
values of $L_{\rm sym}$. We note that these models do allow for maximum
masses in the range 
$2.0\,M_{\odot} \lesssim M_{\rm max} \lesssim 2.3\,M_{\odot}$.

Another important point of our analysis is that our EoS models
with or without heavy baryons which predict very similar
\MR relations have different compositions with different sets of
neutrino emission processes and, therefore, cooling behavior. 
The core matter of nucleonic CS does not allow for the direct Urca
process, as the low values of coefficient $L_{\rm sym}$ used in our
models result in a low proton fraction. Note that the DDME2
parametrization, i.e., without variations of the
$L_{\rm sym}$ parameter does not allow direct Urca as well.
In the case of the $\Delta$-admixed hyperonic matter, direct Urca processes
involving $\Delta$'s and hyperons are allowed for a large region of
the density relevant for high mass CSs. Finally, we showed
that the presence of heavy baryons in CSs does not alter the
universalities of the $I$-Love-$Q$ relations for static
configurations. We further extended this relation to the low-mass
domain not been studied previously, but which is relevant for the 
CCO in HESS J1731-347. These relations were found accurate 
up to 1\% for mass range from $0.2$-$2.5\,M_{\odot}$.

\nolinenumbers
\section*{Acknowledgements}
J.~L. acknowledges the support of the NNSF 
of China (No. 12105232), the Fundamental Research Funds for the Central 
Universities (No. SWU-020021), and by the Venture \& Innovation Support 
Program for Chongqing Overseas Returnees (No. CX2021007).
A.~S. is supported by the Deutsche Forschungsgemeinschaft Grant 
No. SE 1836/5-2 and the Polish NCN Grant
No. 2020/37/B/ST9/01937 at Wroc\l{}aw University.

%\section*{Declaration of competing interest}
%The authors declare that they have no known competing financial interests
%or personal relationships that could have appeared to influence the work
%reported in this paper.

%\section*{Data availability}
%Data will be made available on request.

%\bibliographystyle{elsarticle-num}
%\bibliography{Hess_refs}

\end{document}